%% file: main.tex
\PassOptionsToPackage{hyphens}{url}

\documentclass[journal,preprint]{vgtc}                     

\onlineid{1998}

\preprinttext{To appear in IEEE Transactions on Visualization and Computer Graphics.}

\vgtccategory{Research}
\vgtcpapertype{algorithm/technique}

\input{pages/Xcommands}

\title{Privacy-Preserving Gaze Data Streaming in Immersive Interactive Virtual Reality: Robustness and User Experience}

\author{%
  \authororcid{Ethan Wilson}{0000-0003-0944-2641},
  \authororcid{Azim Ibragimov}{0009-0002-4325-9277}, \authororcid{Michael J. Proulx}{0000-0003-4066-3645}, Sai Deep Tetali, \authororcid{Kevin Butler}{0000-0002-7498-4239}, and \authororcid{Eakta Jain}{0000-0001-5131-3355}
}

\authorfooter{
  \item
  	Ethan Wilson, Azim Ibragimov, Kevin Butler, and Eakta Jain are with the Computer \& Information Science \& Engineering department at the University of Florida.  E-mail: \{ethanwilson,a.ibragimov,butler,ejain\}@ufl.edu.
   \item
  	Michael J. Proulx is with Meta Reality Labs Research.  E-mail: michaelproulx@meta.com.
  \item
  	Sai Deep Tetali is with Meta Reality Labs.  E-mail: saideept@meta.com.
}

\input{pages/0abstract}


\teaser{
  \centering
    \includegraphics[width=\linewidth]{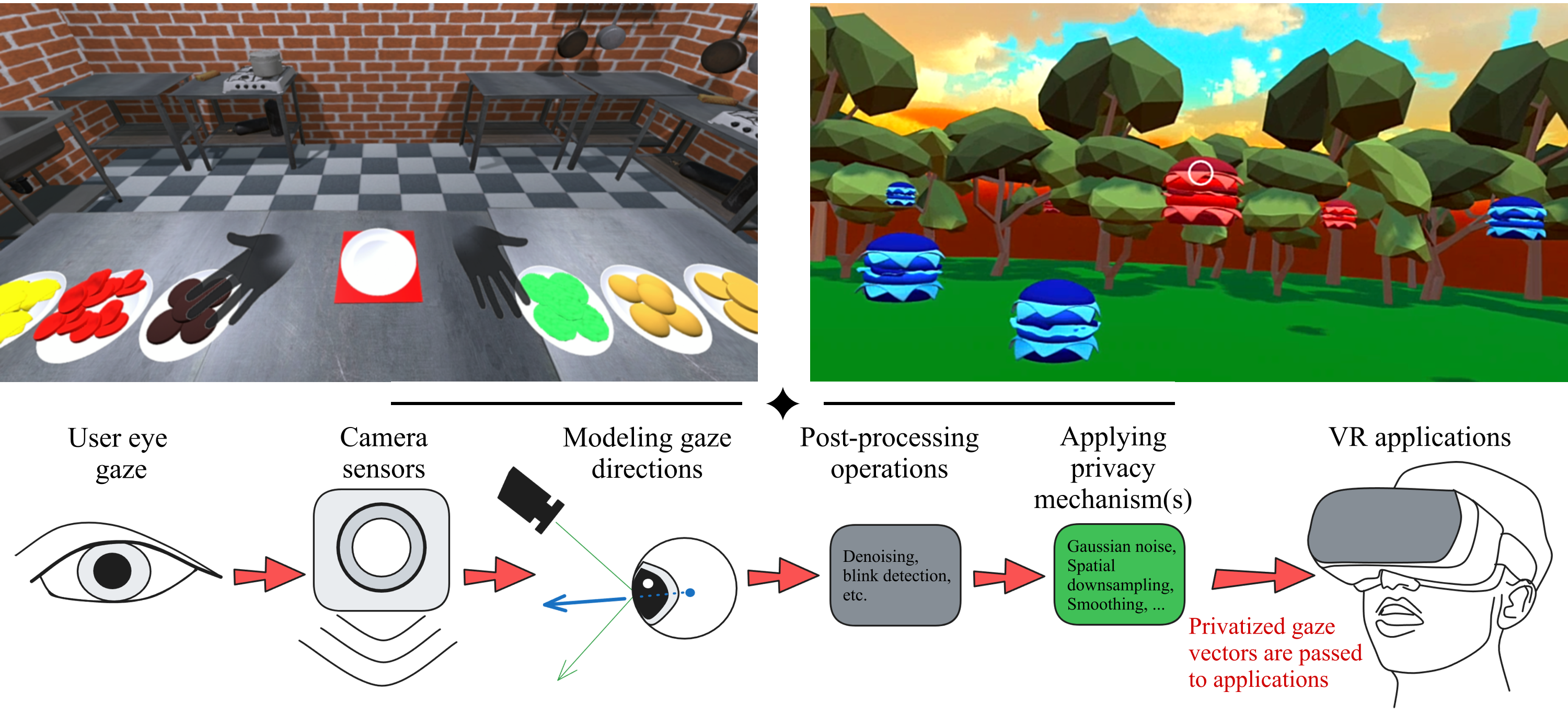}
  \caption{
  \editadd{[R3] How to deploy such a real-time defense? I assume this is a software defense, which means it requires no additional hardware. It is unclear to me how to deploy the defense module in an AR/VR headset or within its firmware. Is it running on the same core processor? I expect to see more discussions on this aspect.}
  {\editnote{We updated the teaser figure to better illustrate where in the pipeline our privacy mechanisms can be applied. 
 At the same time as other post-processing operations on calculated gaze angles (the exact operations are proprietary, but may include smoothing to remove jitter, blink detection, averaging between the two eyes, etc.), we can apply our privacy mechanisms securely then make only those private gaze angles visible to applications.} \textbf{Top-left:} Participants underwent an interactive VR experience while eye movements were passively recorded; \textbf{Top-right:} Participants played an eye tracked VR game with privacy mechanisms applied to gaze streams; \textbf{Bottom:} An illustration of the eye tracking pipeline in head-mounted virtual reality systems.  After eye images are processed and gaze vectors are calculated, we can apply privacy mechanisms to gaze vectors securely on the headset \textit{before} passing values to potentially untrustworthy applications.}}
  \label{fig:teaser}
}

\graphicspath{{figs/}{figures/}{pictures/}{images/}{./}} 

\usepackage{tabu}                      
\usepackage{booktabs}                  
\usepackage{lipsum}                    
\usepackage{mwe}                       

\usepackage{mathptmx}                  

\begin{document}


\firstsection{Introduction}

\maketitle

\input{pages/1introduction}
\input{pages/2related-work}
\input{pages/3-4preliminary-experiment}
\input{pages/5user-in-the-loop}
\input{pages/6threat-scenarios}
\input{pages/7discussion}
\input{pages/8conclusion}

\acknowledgments{
We acknowledge funding support from NSF Award CNS-2206950 and a Privacy-Enhancing Technology Award from Meta.
}

\bibliographystyle{abbrv-doi-hyperref}
\bibliography{references,misc_references}

\input{pages/zappendix}

\end{document}

%% file: pages/Xcommands.tex
\usepackage{lipsum}
\usepackage{gensymb}
\usepackage{caption}
\usepackage{subcaption}
\usepackage{enumitem}
\usepackage[normalem]{ulem}
\usepackage{algpseudocode}
\usepackage{algorithm}
\usepackage{soul}
\usepackage{changepage}   
\usepackage{amsmath}
\DeclareMathOperator*{\argmax}{arg\!max}
\DeclareMathOperator*{\argmin}{arg\!min}
\usepackage{color}
\usepackage{tabularray}
\usepackage{booktabs}

\usepackage{multirow}
\usepackage{colortbl}
\usepackage{hhline}
\usepackage{arydshln}
\usepackage{comment}
\usepackage[pagebackref,bookmarks]{hyperref}

\usepackage{multirow}

\sethlcolor{yellow}

\newenvironment{indentparagraph}{\begin{adjustwidth}{0.5cm}{}}{\end{adjustwidth}}

\def\revisionmode{0} 

\newcommand{\editadd}[2]{%
\if\revisionmode1%
    \begin{color}{blue}%
    \noindent\rule[0.5ex]{\linewidth}{1pt}%
    \textbf{Reviewer Comment: #1}%
    \\%
    #2%
    \noindent\rule[0.5ex]{\linewidth}{1pt}%
    \end{color}%
\else%
#2%
\fi%
}%

\newcommand{\editaddsmall}[1]{%
\if\revisionmode1%
    \textcolor{blue}{#1}%
\else%
#1%
\fi%
}%

\newcommand{\editremove}[1]{%
\if\revisionmode1%
    \textcolor{red}{\sout{#1}}%
\fi%
}%

\newcommand{\editnote}[1]{%
\if\revisionmode1%
    \hl{#1}%
\fi%
}%


%% file: pages/0abstract.tex
\abstract{Eye tracking is routinely being incorporated into virtual reality (VR) systems. 
Prior research has shown that eye tracking data, if exposed, can be used for re-identification attacks~\cite{david-john_privacy-preserving_2021}. 
The state of our knowledge about currently existing privacy mechanisms is limited to privacy-utility trade-off curves based on data-centric metrics of utility, such as prediction error, and black-box threat models.
We propose that for interactive VR applications, it is essential to consider user-centric notions of utility and a variety of threat models.
We develop a methodology to evaluate real-time privacy mechanisms for interactive VR applications that incorporate subjective user experience and task performance metrics.  
We evaluate selected privacy mechanisms using this methodology and find that re-identification accuracy can be decreased to as low as 14\% while maintaining a high usability score and reasonable task performance.
Finally, we elucidate three threat scenarios (black-box, black-box with exemplars, and white-box) and assess how well the different privacy mechanisms hold up to these adversarial scenarios. 
This work advances the state of the art in VR privacy by providing a methodology for end-to-end assessment of the risk of re-identification attacks and potential mitigating solutions.

\if\revisionmode1

\textbf{REVISION LEGEND:}

\textcolor{blue}{Additions are colored blue.  For larger additions in response to the meta review or reviewer feedback, we have included the comment in bold text above the section.}

\textcolor{red}{Removed text is written in red with strikethrough.}

\hl{Some notes have been written, either to explain other changes or to elaborate on our responses to reviewer feedback.  These are highlighted, and this text does not appear in the final version at all.}

\fi

}

%% file: pages/1introduction.tex

Virtual reality (VR) technology has seen a rapid deployment of eye tracking-enabled headsets over the past several years.  Eye tracking has many applications in VR, including as an interaction modality~\cite{monteiro_hands-free_2021,fernandes_leveling_2023, piumsomboon_exploring_2017, gowases_gaze_2008, atienza_interaction_2016, orlov_effectiveness_2015}, as an animation tool~\cite{seele_heres_2017, normoyle_evaluating_2013, ruhland_perception_2015}, for attention analysis~\cite{wang_how_2021, liu_impacts_2022}, for rendering optimizations~\cite{lungaro_gaze-aware_2018, weier_foveated_2016, weier_foveated_2018, kaplanyan_deepfovea_2019, david-john_towards_2021, xu_gaze_2018, hu_dgaze_2020}, and for user authentication~\cite{makowski_deepeyedentificationlive_2021, lohr_eye_2022-1, lohr_eye_2022, jia_biometric_2018, peng_eye_2022}.

Recently, eye tracking movements alone have been found to function as a biometric identifier~\cite{kasprowski_eye_2004, george_score_2016}.  Users can be uniquely identified in small directories; in VR, users have been shown to be identifiable at up to 85\% accuracy~\cite{lohr_eye_2022}.  
\editadd{[Meta Review] The motivations should be further clarified and enhanced.}{This opens the risk of unwanted re-identification in online VR usage.  Eye gaze is a promising input device showcasing very unique interactions and optimizations, but users should not have to choose between these interactions or their own privacy.}
If eye tracking data is shared with the proper safeguards, the risk of re-identification attacks or unwanted data leakage for VR users is alleviated.


Some solutions have been proposed to protect users against re-identification while still enabling eye tracking data utility.  Existing analyses of these mechanisms focus on data-centric utility, including downstream processes such as area of interest analysis or gaze prediction.  However, analyses focusing on interactive VR need to consider the user, who is likely to feel the effects when their eye tracking data streams are perturbed to grant privacy.  It is necessary to consider the user first when designing privacy mechanisms that will be incorporated into interactive VR experiences.  Users may chafe at adopting privacy solutions that impact the interactive experience (which is the reason for being in VR in the first place).  Another limitation to current eye tracking privacy methodologies is that they are framed as black-box threat scenarios.  In real world settings, however, malicious 3rd parties have access to a number of strategies that aim to nullify the discussed privacy efforts.

\input{pages/2.table}

\textbf{Our contributions:}
We first update current knowledge on re-identification risk in interactive VR by using the state-of-the-art in eye movement identity matching and developing an interactive VR experience to serve as an evaluation testbed, re-identifying $N=26$ users at an average accuracy of 67.3\%.  
We \editremove{then} measure the privacy capabilities of multiple mechanisms, identifying the following as capable of protecting privacy: Gaussian noise, spatial downsampling, and smoothing.  Next, we incorporate these mechanisms into an interactive VR experience which uses eye tracking as the primary mode of control.  We jointly measure re-identification accuracy, task performance, and qualitative subjective utility responses to firmly assess the privacy-utility trade-off of these mechanisms.  Our mechanisms decrease re-identification rates as low as 14.1\% while retaining high subjective usability and reasonable task performance.  Finally, we evaluate the mechanisms' robustness against dedicated adversaries under three plausible threat scenarios: black-box access, black-box access with exemplars, and white-box access.  Our provided methodology can serve as a guide for future research of privacy mechanisms in interactive settings, measuring along multiple axes. 
The evaluated privacy mechanisms can be utilized as a basis for further innovation of novel privacy mechanisms.

\textbf{Broader Impacts:}
This research contributes to the advancement of eye tracking privacy mechanisms, specifically those that must be applied to sample-level data in real-time.  These mechanisms protect users against detection, especially populations that could be vulnerable if they are identified.  Our work highlights a necessary shift in focus for the virtual reality research community from data-driven notions of utility to a user-centric design~\cite{miller_personal_2020}.  We additionally highlight that a calculated re-identification rate is only the first step; in situations where formal privacy guarantees can not be reached, we must consider real-world threat scenarios in order to proactively protect against adversaries.  \editadd{[R3] I highly recommend the authors make their dataset publicly available to other scholars to advance future research.}{In addition, we make our collected datasets containing eye tracking data in interactive VR scenes available at \url{https://doi.org/10.5281/zenodo.10475455}.}


%% file: pages/2.table.tex
\begin{table*}[t]
\small
\centering
\resizebox{\linewidth}{!}{%
\begin{tabular}{l|l|l|l|c|l|c} 
\toprule
\textbf{Application}                                                             & \textbf{Data Domain}                                                                & \textbf{Data Format}                                      & \textbf{Mechanism}                                                                                                                                    & \multicolumn{1}{l|}{\textbf{ID Accuracy}} & \textbf{Data-centric Utility}                                                            & \multicolumn{1}{l}{\textbf{User-centric Utility}}                                                                                                                               \\ 
\hline
\multirow{11}{*}{\begin{tabular}[c]{@{}l@{}}Gaze-based\\analytics\end{tabular}}  & \multirow{7}{*}{Constrained VR}                                                     & Sample-level                                              & \begin{tabular}[c]{@{}l@{}}(Kaleido) Spatial noise \\with adaptive sampling~\cite{david-john_privacy-preserving_2023}\end{tabular} & 28\% to 6\%                                              & Activity classification                                                                  & -                                                                                                                                                                               \\ 
\cline{3-7}
                                                                                 &                                                                                     & Sample-level                                              & Gaussian noise~\cite{david-john_privacy-preserving_2021}                                                                           & 85\% to 30\%                                             & Dwell time RMSE                                                                          & -                                                                                                                                                                               \\ 
\cline{3-7}
                                                                                 &                                                                                     & Sample-level                                              & Temporal downsampling~\cite{david-john_privacy-preserving_2021}                                                                    & 85\% to 79\%                                             & Dwell time RMSE                                                                          & -                                                                                                                                                                               \\ 
\cline{3-7}
                                                                                 &                                                                                     & Sample-level                                              & Spatial downsampling~\cite{david-john_privacy-preserving_2021}                                                                     & 85\% to 48\%                                             & Dwell time RMSE                                                                          & -                                                                                                                                                                               \\ 
\cline{3-7}
                                                                                 &                                                                                     & Sample-level                                              & Gaussian noise~\cite{david-john_privacy-preserving_2021}                                                                           & 33\% to 9\%                                              & KL-divergence of saliency maps                                                           & -                                                                                                                                                                               \\ 
\cline{3-7}
                                                                                 &                                                                                     & Sample-level                                              & Temporal downsampling~\cite{david-john_privacy-preserving_2021}                                                                    & 9\% to 7\%                                               & KL-divergence of saliency maps                                                           & -                                                                                                                                                                               \\ 
\cline{3-7}
                                                                                 &                                                                                     & Sample-level                                              & Spatial downsampling~\cite{david-john_privacy-preserving_2021}                                                                     & 47\% to 29\%                                             & KL-divergence of saliency maps                                                           & -                                                                                                                                                                               \\ 
\cline{2-7}
                                                                                 & \multirow{4}{*}{\begin{tabular}[c]{@{}l@{}}Conventional\\eye tracking\end{tabular}} & \begin{tabular}[c]{@{}l@{}}Aggregate-\\level\end{tabular} & \begin{tabular}[c]{@{}l@{}}Difference- and chunk-based \\Fourier perturbation~\cite{bozkir_differential_2021}\end{tabular}         & 100\% to 28\%                                            & \begin{tabular}[c]{@{}l@{}}Document classification; \\gender classification\end{tabular} & -                                                                                                                                                                               \\ 
\cline{3-7}
                                                                                 &                                                                                     & \begin{tabular}[c]{@{}l@{}}Aggregate-\\level\end{tabular} & \begin{tabular}[c]{@{}l@{}}Exponential mechanism\\applied to features~\cite{steil_privacy-aware_2019}\end{tabular}                 & 100\% to $\sim$10\%                           & \begin{tabular}[c]{@{}l@{}}Document classification; \\gender classification\end{tabular} & -                                                                                                                                                                               \\ 
\cline{3-7}
                                                                                 &                                                                                     & \begin{tabular}[c]{@{}l@{}}Aggregate-\\level\end{tabular} & k-same-synth~\cite{david-john_privacy-preserving_2023}                                                                             & 28\% to 7.5\%                                            & Activity classification                                                                  & -                                                                                                                                                                               \\ 
\cline{3-7}
                                                                                 &                                                                                     & \begin{tabular}[c]{@{}l@{}}Aggregate-\\level\end{tabular} & \begin{tabular}[c]{@{}l@{}}Event-synth-\\plausible deniablity~\cite{david-john_privacy-preserving_2023}\end{tabular}               & 28\% to 14.2\%                                           & Activity classification                                                                  & -                                                                                                                                                                               \\ 
\hline
\multirow{2}{*}{\begin{tabular}[c]{@{}l@{}}Gaze-based\\interaction\end{tabular}} & \begin{tabular}[c]{@{}l@{}}Webcam\\eye tracking\end{tabular}                        & Sample-level                                              & \begin{tabular}[c]{@{}l@{}}(Kaleido) Spatial noise \\with adaptive sampling~\cite{li_kalido_2021}\end{tabular}                     & $\sim$84\% to $\sim$8\%            & \begin{tabular}[c]{@{}l@{}}scan path similarity; \\latency trade-off\end{tabular}        & \multicolumn{1}{l}{\begin{tabular}[c]{@{}l@{}}Game enjoyment;\\task performance\end{tabular}}                                                                                  \\ 
\hhline{~------}\cline{2-7}
                                                                                 & {\cellcolor[rgb]{0.902,0.902,0.902}}Interactive VR                                  & {\cellcolor[rgb]{0.902,0.902,0.902}}Sample-level          & {\cellcolor[rgb]{0.902,0.902,0.902}}\textbf{Ours}                                                                                                              & {\cellcolor[rgb]{0.902,0.902,0.902}}67.3\% to 14.1\%     & {\cellcolor[rgb]{0.902,0.902,0.902}}Area of interest retention                         & \multicolumn{1}{l}{{\cellcolor[rgb]{0.902,0.902,0.902}}\begin{tabular}[c]{@{}>{\cellcolor[rgb]{0.902,0.902,0.902}}l@{}}Subjective usability; \\task performance\end{tabular}} 
\end{tabular}
}
\caption{Collection of eye tracking privacy work that successfully protected against re-identification while retaining one or more measure(s) of utility.}
\label{tab:related-work}
\end{table*}

%% file: pages/2related-work.tex
\section{Related Work}

Eye tracking is becoming a prominent feature of VR experiences.  Research on eye tracked VR systems began more than two decades ago~\cite{tanriverdi_interacting_2000, duchowski_binocular_2000}.  In recent years, many commercial VR head-mounted displays (HMDs) have released with embedded eye trackers (Magic Leap 1~\cite{magic_leap_1} in 2018, Vive Pro Eye~\cite{vive_pro_eye} in 2019, HoloLens 2~\cite{hololens_2} in 2019, Vive Focus 3~\cite{vive_focus_3} in 2021, Magic Leap 2~\cite{magic_leap_2} in 2022, and Meta Quest Pro~\cite{meta_quest_pro} in 2023.  The Apple Vision Pro~\cite{apple_vision_pro} is set to release in early 2024).  These hardware advancements have created a surge in interest at the intersection between eye tracking and VR.  

\subsection{Applications of Eye tracking data in VR}

Eye tracking enables many interactions in virtual scenes.  These include using gaze to directly interact with virtual objects, improving social VR interactions, enabling foveated rendering optimizations, and gaze analysis as a research tool.

\textbf{Gaze-based interaction:}
Eye tracking movements allow users to interact with virtual scenes, either on their own or paired with other control modes~\cite{monteiro_hands-free_2021,fernandes_leveling_2023}.  Gaze direction can aim a cursor along with button presses to select objects~\cite{piumsomboon_exploring_2017, gowases_gaze_2008}, or gaze fixations can be used for selection~\cite{gowases_gaze_2008, atienza_interaction_2016}.  \editadd{[R1] The authors may want to make this point clearer at the beginning of the paper. and introduce more related applications in the related work.}{For example, in a VR application interactable objects may glow when looked at, to indicate that they are dynamic.  Then by fixating while pressing a button, users can select these objects\footnote{\url{https://www.uploadvr.com/polyarc-moss-psvr-2/}}.}  The upcoming Apple Vision Pro will support gaze-controlled interfaces paired with pinching gestures~\cite{apple_vision_pro}.  Gaze-based interactions in desktop games are found to be more efficient and immersive than traditional control modes~\cite{gowases_gaze_2008, orlov_effectiveness_2015}, which may persist into VR\footnote{A list of VR games which incorporate eye tracking: \url{https://www.psfanatic.com/here-are-all-the-psvr2-games-that-use-eye-tracking-in-cool-ways/}}.

\textbf{Rendering Optimization:}
Foveated rendering is a critical optimization to increase resolution and frame rate of VR headsets~\cite{guenter_foveated_2012, lungaro_gaze-aware_2018}.  Foveated rendering sparsely renders samples outside of the fovea region, which is determined  through eye tracking signals.  Because peripheral vision has lower acuity than foveal vision, the image could be perceptually similar to traditional rendering but vastly less expensive to compute~\cite{weier_foveated_2016, weier_foveated_2018, kaplanyan_deepfovea_2019}.  Gaze prediction algorithms~\cite{david-john_towards_2021, xu_gaze_2018, hu_dgaze_2020} will enable proactive foveated rendering and occlusion optimizations.

\textbf{Avatar Animation:}
Recorded gaze can be used to drive eye animations in VR.  More realistic eye movements have been shown in improve the quality of interaction with virtual avatars~\cite{garau_impact_2003} and to increase perceived presence and avatar realism~\cite{seele_heres_2017}.  Gaze can enable virtual avatars to display trust across multiple expressions and contexts~\cite{normoyle_evaluating_2013}, and multiple personality traits can be discerned solely through characters' eye motions~\cite{ruhland_perception_2015}.  By incorporating real gaze behaviors to embodied avatars, each avatar feels more unique and personable.

\textbf{Gaze-based Analytics:}
Eye tracking data can be a rich tool for data analysis~\cite{shadiev_review_2023}.  Examples include area of interest (AOI) calculation~\cite{david-john_privacy-preserving_2021, orquin_areas_2016}, document classification and analysis~\cite{david-john_for_2022, sanches_using_2017}, and attention visualization~\cite{liu_differential_2019}.  Researchers from multiple fields use eye movements to analyze topics such as social behavior~\cite{reichenberger_gaze_2020, zito_street_2015}, visual attention~\cite{wang_how_2021, liu_impacts_2022}, and simulated responses under stress~\cite{shi_neurophysiological_2020}.


\subsection{Identification Risk of Recorded Eye Movements}

Though users can be identified based on several cues, including head movements, body movements~\cite{miller_temporal_2022} and gestures~\cite{miller_combining_2022}, our focus is identification based on eye movements. Iris patterns are a well known biometric identifier~\cite{negin_iris_2000}, and David-John et al. examined user re-identification using iris images and presented solutions to mitigate this risk~\cite{john_let_2020, john_security-utility_2020}.  Note that the Meta Quest Pro headsets used in this study do not pass on eye images or raw data to the applications.  In addition to hand-crafted features derived from gaze streams~\cite{kasprowski_eye_2004, schroder_robustness_2020, george_score_2016, lohr_eye_2020}, there are now deep-learning methods to classify users based on short windows of eye movement data~\cite{makowski_deepeyedentificationlive_2021, lohr_eye_2022-1, lohr_eye_2022, jia_biometric_2018, peng_eye_2022}.  Eye Know You Too (EKYT) is currently the top performing eye movement identification model, reporting accuracies as high as 91.38\% on 1000Hz data~\cite{lohr_eye_2022}. 

Physical and behavioral attributes such as personality~\cite{berkovsky_detecting_2019}, age~\cite{zhang_how_2018} or gender~\cite{sammaknejad_gender_2017} have been inferred from eye movements.  Some research leverages eye movements to aid in medical diagnoses such as Autism or Alzheimer's~\cite{vargas-cuentas_developing_2017, wan_applying_2019, sun_novel_2022}.  While there are appropriate use cases to learn this information, users can not consciously hide the attributes embedded within eye tracking streams.  The opportunity here is develop methods to block these features from being extracted without users' consent by malicious entities who acquire eye tracking data.  

The threat of eye movement re-identification is larger for small sets of users~\cite{friedman_biometric_2022}.  This may be a particular concern for marginalized users who face disproportionate harm when privacy is compromised~\cite{sannon_privacy_2022}.

\subsection{Gaze Data Privacy}

Privacy mechanisms are mainly applied in three ways to the eye tracking pipeline. Aggregate-level mechanisms protect full datasets with operations that average across multiple users' data~\cite{liu_differential_2019, david-john_privacy-preserving_2023, bozkir_differential_2021, steil_privacy-aware_2019}.  Feature-level mechanisms protect users by converting raw gaze signals to features and applying privacy~\cite{david-john_for_2022, david-john_privacy-preserving_2023, bozkir_differential_2021, steil_privacy-aware_2019}.  Sample-level mechanisms operate on the actual data streams, perturbing gaze direction at every frame~\cite{david-john_privacy-preserving_2021, li_kalido_2021}.  In VR, we are mainly interested in sample-level mechanisms~\cite{selinger_eye-tracking_2023}, which could be applied securely by the VR platform before eye tracking data is made available to third party applications~\cite{david-john_privacy-preserving_2021}.  \editadd{[Meta Review ]The motivations should be further clarified and enhanced.}{With privacy mechanisms in place, users can experience novel interactions and optimizations only possible with sample-level gaze streams without risking information leakage.}  See Table~\ref{tab:related-work} for a collection of eye tracking privacy work.  

\editadd{[R2] The authors should have been more clear about their advancements on the field, since it's the main contribution and focus of this work.}{Real-time privacy operations will be critical to ensure privacy during online eye tracking interactions~\cite{piumsomboon_exploring_2017, atienza_interaction_2016} and to enable optimizations~\cite{lungaro_gaze-aware_2018}, but have yet to be explored in real-time VR settings.}  The goal of privatization is to protect sensitive attributes while keeping the data usable with respect to a given task.  Existing research primarily measures data-driven utility via post-processing tasks, such as gaze-based analytics~\cite{liu_differential_2019, bozkir_differential_2021, david-john_for_2022} or rendering optimization~\cite{david-john_towards_2021}.  However, the impact of privacy mechanisms on the user's performance and subjective experience in interactive VR has not been considered prior to this work.

%% file: pages/3-4preliminary-experiment.tex
\section{Re-identification in Interactive VR}


We begin by establishing the risk of re-identification in interactive VR and identifying viable privacy mechanisms.  We collect a dataset of eye tracking-enabled interactive VR tasks to serve as an evaluation testbed.  We then quantify the re-identification risk on our dataset using the state of the art architecture and define and evaluate multiple privacy mechanisms on our dataset.

On conventional eye trackers at high frequencies, users can be identified at very high accuracies (91.38\% identification rate, 3.66\% equal error rate)~\cite{lohr_eye_2022}.  In VR settings, it is less clear how reliably users can be identified, due to less precise sensors and extraneous user movements.  An evaluation of 360\textdegree \space VR image and video datasets using a prior identification method~\cite{george_score_2016} yielded identification rates ranging from 9\% to 85\%~\cite{david-john_privacy-preserving_2021}.  The same analysis also evaluated an interactive dataset where users viewed a scene of moving animals~\cite{hu_dgaze_2020}, yielding only a 3\% identification rate.  The level of interactivity and amount of user movement in VR setups could negatively correlate with the potential to be identified in VR~\cite{ugwitz_eye-tracking_2022}.  In this paper, we present an up-to-date evaluation of re-identification risk in interactive VR tasks using the state-of-the-art identification model.



For this analysis, we distinguish eye tracking data sources into three categories.  \textit{Conventional eye tracking} utilizes high quality static sensors at 1000Hz or greater~\cite{griffith_gazebase_2021}, and produces highly identifiable data~\cite{lohr_eye_2022}. Eye movements collected in VR can be separated into two categories: \textit{Interactive VR} is presented in a natural way.  Users directly interact with dynamic objects in the virtual scene, and different users experience the scene at their own rate.  This is reflective of consumer applications, such as VR games or dynamic training scenarios.  On the other hand, \textit{constrained VR} is representative of existing experimental setups.  User movements are limited, such as sitting in a chair~\cite{sitzmann_saliency_2018} or placing the head on a chinrest~\cite{lohr_gazebasevr_2023}, and the tasks are standardized such that all users experience the same stimuli at the same rates.  
In this section, we answer the following research questions:

\begin{itemize}[noitemsep, topsep=0pt]
    \item How do current state of the art eye movement re-identification algorithms perform in interactive VR?
    \item What real-time privacy mechanisms are effective at protecting against re-identification in interactive VR tasks?
\end{itemize}

We present an updated evaluation of the risk of identification in VR.  Prior work only considered identification risk on constrained VR setups using models trained on hand-crafted features~\cite{david-john_privacy-preserving_2021}.  We first construct a dynamic VR game representative of interactive VR.  We then present an analysis of the current state of the art architecture trained both on conventional eye tracking data and VR data, evaluating on conventional eye tracking data (GazeBase~\cite{griffith_gazebase_2021}), constrained VR data (GazeBaseVR~\cite{lohr_gazebasevr_2023}), and interactive VR data (our dataset).  We then discuss our results, giving insights to the current risk of identification in consumer VR and the relationship between re-identification potential and the amount of data made available.  Using the same dataset, we then evaluate multiple privacy mechanisms that can be applied to protect eye tracking data in VR.  We evaluate these mechanisms across increasing intensities to derive well saturated privacy curves.

\subsection{Data Collection Methodology}
\label{sec:e1-dataset}

We describe the protocol for our collected dataset which serves as a testbed to evaluate identification in interactive VR.

\textbf{Participants:}
Survey participants were recruited under IRB approved protocol via several communication channels including word of mouth and electronic mailing list advertisements ($N=26$; 57.69\% male, 42.31\% female).  No monetary compensation was provided, but some participants received extra credit for undergraduate courses.  Eligible participants required normal or corrected-to-normal vision without the use of eye glasses.  The racial-ethnic distribution is 61.54\% White, 19.23\% Indian Asian, 15.38\% Black or African American, 7.69\% Eastern Asian, and 7.69\% Hispanic/Latino; 11.54\% of participants report two or more races.  34.62\% of participants were age 18-20, 50\% 21-29, 11.54\% 30-39, and 3.85\% 50-59.  88.46\% of participants reported some level of experience with VR, and 30.77\% reported some experience with using eye tracking as a control mode.

\textbf{Procedure:}
Participants were instructed to act as employees in a sandwich shop, and were tasked with assembling as many sandwiches as possible in a 90 second time frame.  Plates of stacked ingredients were organized on the sandwich assembling counter on each side.  In the center of the counter, participants would assemble their sandwiches on an empty plate (See Figure~\ref{fig:teaser}a).  After a sandwich was completed, a small animation would play and the plate would be cleared, allowing participants to begin the next sandwich.  A digital timer could be seen which displayed the amount of time remaining.  

Before beginning the main task, participants were encouraged to practice picking up and assembling ingredients.  In the experiment, participants did not use controllers; instead, grabbing was driven by hand tracking technology to provide a more immersive experience~\cite{lin_effect_2019}.  Participants underwent 4 trials of the same task and were allowed time to rest between trials.  

As participants performed the tasks, gaze data was collected passively to evaluate the potential of re-identification attacks and to store area of interest (AOI) data.  Each ingredient plate, the assembly plate, and the timer were all AOI regions with fully-covering bounding boxes being used for AOI collision detection.

\textbf{Validation:}
\label{sec:validation}
Data was collected using the Meta Quest Pro~\cite{meta_quest_pro} (1920 x 1800 pixels per eye, 72Hz refresh rate).  Before the experiment, each participant underwent the headset's eye tracking calibration procedure.  Participants gazed directly at spheres which appeared at random points on the screen and gradually shrank until no longer visible.  To validate the accuracy of collected gaze data, participants performed an eye tracking validation task.  Participants faced a $3 \times 3$ checkerboard of red targets situated 2 meters away spanning a $38.58 \degree$ angle vertically and horizontally.  Participants were instructed to gaze directly at the active target, which would become green.  The active target cycled uniformly; each target was active for 2 seconds total, with no downtime until the next dot became active.  We report a spatial accuracy error of $\mu = 2.64\degree$, $\sigma = 1.24$.  This protocol is comparable with recent analysis of the Quest Pro~\cite{wei_preliminary_2023}.

\subsection{Privacy Mechanisms}

Collected gaze streams over a full VR session can be represented as a time series of gaze angles in spherical coordinates $\theta, \psi$ and their corresponding time stamps $t$: $X = \{(\theta_0, \psi_0, t_0),(\theta_1, \psi_1, t_1), \cdots, (\theta_n, \psi_n, t_n)\}$.  The gaze angles are localized relative to the recorded position of the headset, i.e. head pose, thus are constrained roughly to the human field of view.  As our privacy mechanisms are implemented in real-time, the operations are applied directly on the frame which the gaze is sampled.

We implement three privacy protection mechanisms proposed by David-John et al.~\cite{david-john_privacy-preserving_2021}: additive Gaussian noise, temporal downsampling, and spatial downsampling.  We also introduce linearly weighted average smoothing as a novel mechanism.  \editaddsmall{These mechanisms are designed to be feasible in real-time settings, suitable for streamed sample-level eye tracking data in VR applications.}

\textbf{Gaussian Noise:}
Gaussian noise is sampled independently for the $\theta, \psi$ gaze angles for every frame.  We use the standard deviation of the noise sample $\sigma$ to control the privacy-utility trade-off provided by the method, yielding the following per-frame operation: $X'_n = (\theta_n + x \sim \mathcal{N}(0, \sigma), \psi_n + y \sim \mathcal{N}(0, \sigma), t)$.  

\textbf{Temporal Downsampling:}
Temporal downsampling effectively lowers the sampling rate of a stream of data by a factor $K$.  Data entries with indices where $K$ is not a factor are removed from the data stream, yielding a stream with $N/K$ total entries after downsampling.  

A real-time application may expect a gaze vector at all frames.  Thus, we preserve the original data format by simply copying the gaze directions from the prior time step on frames that would traditionally be removed.  Given a downsampling factor $K$, 
\[
X'_n = \left\{ 
  \begin{array}{ c l }
    (\theta_n, \psi_n, t)               & \quad \textrm{if } n \mathbin{\%} K = 0 \\
    (\theta'_{n-1}, \psi'_{n-1}, t)   & \quad \textrm{otherwise}
  \end{array}
\right.
\]

\textbf{Spatial Downsampling:}
Spatial downsampling lowers the spatial resolution of the gaze data, e.g., multiple nearby full-resolution points would be mapped to a single down-sampled point, lowering the spatial fidelity.  To apply spatial downsampling to continuous gaze angles, we first map the angles into a set of discrete points large enough to preserve data quality, choosing 2160 points to cover a $180\degree$ field of view.  We achieve spatial downsampling by remapping the gaze angles into a smaller domain equal to the reference domain divided by $L$.  We map the $\theta, \psi$ angles into the discrete domain of $M = 2160 / L$, providing a step size of $\delta = \frac{180\degree}{M}$.  This yields the following per-frame operation: $X'_n = (\lfloor \theta / \delta \rfloor \cdot \delta, \lfloor \psi / \delta \rfloor \cdot \delta, t)$. 

\textbf{Smoothing:}
We introduce a smoothing operation as a novel privacy protection mechanism.  Smoothing streamed gaze data can remove identifiable features without displacing individual samples in a jarring way as the above mechanisms can.  Because of this, we hypothesize that users will be accepting of smoothing; they can consciously correct for the gaze stream's behavior by fixating at an objects for a second longer, for example.

To operate in real-time, we define the current gaze vector as a linearly weighted average of $B$ preceding samples.  Preceding samples are stored in a sliding window that is updated every frame.  The smoothed value is a weighted average of the $B$ values in the window, with each sample weighed by its index in the buffer. 
\[
X'_n = \frac{X_{n-B} + 2(X_{n+1-B}) + 3(X_{n+2-B}) + \cdots + B(X_n)}{\sum_{\ i}^{\ B} (i)}
\]

Larger window sizes equal a more intense smoothing operation, which is more successful at removing identifying features, but introduces a larger temporal delay between the input and output gaze vectors.  Implementation details are shown in Algorithm~\ref{alg:smoothing}.

\begin{algorithm}[t]
\caption{Linearly weighted average smoothing}\label{alg:smoothing}
\begin{algorithmic}
\State $B \gets$ \textit{size of window}
\State \texttt{window} $\gets Queue()$ \Comment{Note that this Queue pops at the $0^{th}$ index}
\State $D \gets 0$
\For{$i \gets 1$ to $B$} \Comment{Initializing the window}
    \State \texttt{window}$.add([0, 0])$
    \State $D \gets D + i$
\EndFor 
\While{\textit{application is running}}
    \State $X \gets$ \textit{current gaze vector}
    \State \texttt{window}$.pop()$ 
    \State \texttt{window}$.add(X)$
    \State $X' = (0, 0)$
    \For{$i \gets 1$ to $B$}
        \State $X'[0] \gets X'[0] + $\texttt{window}$[i][0] * i$
        \State $X'[1] \gets X'[1] + $\texttt{window}$[i][1] * i$
    \EndFor
    \State $X' \gets X' / D$ 
\EndWhile
\end{algorithmic}
\end{algorithm}

\subsection{Evaluation Methodology}
\label{sec:privacy-methodology}


We evaluate the identification potential of EKYT models trained on conventional eye tracking data (GazeBase~\cite{griffith_gazebase_2021}) and constrained VR data (GazeBaseVR~\cite{lohr_gazebasevr_2023}).  Both models are trained at 125Hz to best match the frequency of our collected data.  We follow the training and testing methodology of Lohr et al.\footnote{\url{https://dataverse.tdl.org/dataset.xhtml?persistentId=doi:10.18738/T8/61ZGZN}}~\cite{lohr_eye_2022}, and present the rank-1 identity retrieval rate averaged across all tasks in the dataset.

To our knowledge, there is no existing interactive VR dataset at the scale required to train the EKYT model effectively.  Therefore, we assess the effectiveness of training on constrained VR data and evaluating on interactive VR data, versus using a model solely trained on conventional eye tracking data.

To evaluate the identification potential on our collected data, we define the following protocol.  First, embeddings are generated from the raw eye movement data.  Each 90-second trial of the game is separated into 5-second segments sliding over a 1-second interval.  These 5-second segments are linearly interpolated to a constant 125Hz and processed by the EKYT model to create $512D$ feature embeddings $Emb$, which are stored along with labels $L$ for the individuals as records.

Records are separated into folds according to the trials of the game.  So, all records corresponding to the first trial exist in the first fold, and so on.  We then compare pair-wise each fold as a distinct query and reference set.  For each individual, all query records are compared with the records in the embedding set.  So, for $N$ individuals and $M$ records per individuals, $Query_n = \{Emb_{n, 1}, Emb_{n, 2},$ ... $Emb_{n, m}\}$ and reference $Ref = \{\hat{Emb}_{n, m} : \forall n \in N, \forall m \in M \}$.  The record from the query is matched to the closest reference embedding using cosine similarity as a distance metric and the associated label is stored. For an individual $n$, the predicted label $L_{p, n}$ is:

\begin{equation}
\begin{gathered}
    L_{p, n} = \argmax_L \left( \sum_m^M L_{z}\right) \\ \text{where} \quad z = \argmin_m \left( \sum_{n'}^N \sum_{m'}^M \left( 1 - \frac{Query_{n, m} \ \cdot \ Ref_{n', m'}}{||Query_{n, m}|| \ ||Ref_{n', m'}||} \right) \right)
\end{gathered}
\end{equation}

The returned labels are aggregated between each query embedding $Query_{n, m}$ and all embeddings in the reference set to determine a final prediction $L_{p, n}$.  If $L_{p, n}$ equals the true label $L_{n}$, the individual has been successfully identified.

The reported metrics are an average over all individuals and all pair-wise combinations of distinct trials, readable as an average identification accuracy over the full dataset, or the likelihood of any individual being successfully identified.

When evaluating privacy mechanisms, before processing the set of trials which make up the query set, the mechanism is applied at a given strength.  The privatized results are then processed and compared to the reference set.  This represents the following threat scenario:  

\begin{indentparagraph}
    \textit{An adversary acquires a privatized data record without knowing the identity, then queries against a dataset of records which have known identities attached.}
\end{indentparagraph}

To provide an initial data-centric representation of utility, we measure AOI intersections before and after privatization.  We calculate the multi-class weighted precision and recall and overall F1 score which can be interpreted as the mechanism's ability to retain original AOI behavior and after privatization.

\subsection{Results}

\input{pages/3.table}

We organize our results according to the initial analysis of EKYT models on across conventional and VR data domains and the performance of privacy mechanisms on our collected dataset.

\subsubsection{Re-identification Capability of EKYT Models}

Our findings are reported in Table~\ref{tab:tab}.  We report the accuracies between the 2 training and 3 test setups.  We additionally report results varying the duration of each data record.  These results are calculated using the explained methodology but first limiting the amount of data to the first $N$ seconds per data record.

We find that the model trained on conventional eye tracking data performs well when evaluated in the same domain, reaching identification accuracies of 66.09\% and 90.78\% on 5 second and 60 second data records, respectively.  However, the conventional model performs much less effectively in when applied to VR, achieving 27.04\%@5s and 49.98\%@60s on constrained VR and 22.44\%@5s and 56.73\%@60s on interactive VR data.  

On the other hand, the constrained VR model favors VR data at a lower duration (18.04\%@5s for conventional, 33.87\%@5s for constrained VR, 26.92\%@5s for interactive VR) but performs roughly equal on all domains at higher duration (58.57\%@60s for conventional, 55.09\%@60s for constrained VR, 54.81\%@60s for interactive VR).  

We find performance on interactive VR data to be roughly equal between the two models.  While the conventional eye tracking model is trained on a larger dataset, the constrained VR model has closer spatial precision and more similar setup to our data.  For the rest of this paper, we use the constrained VR EKYT model to compute identification accuracies.  

\subsubsection{Privacy Mechanism Performance}
\label{sec:e1-results}

\begin{figure}[t]
    \centering
    \includegraphics[width=1\linewidth]{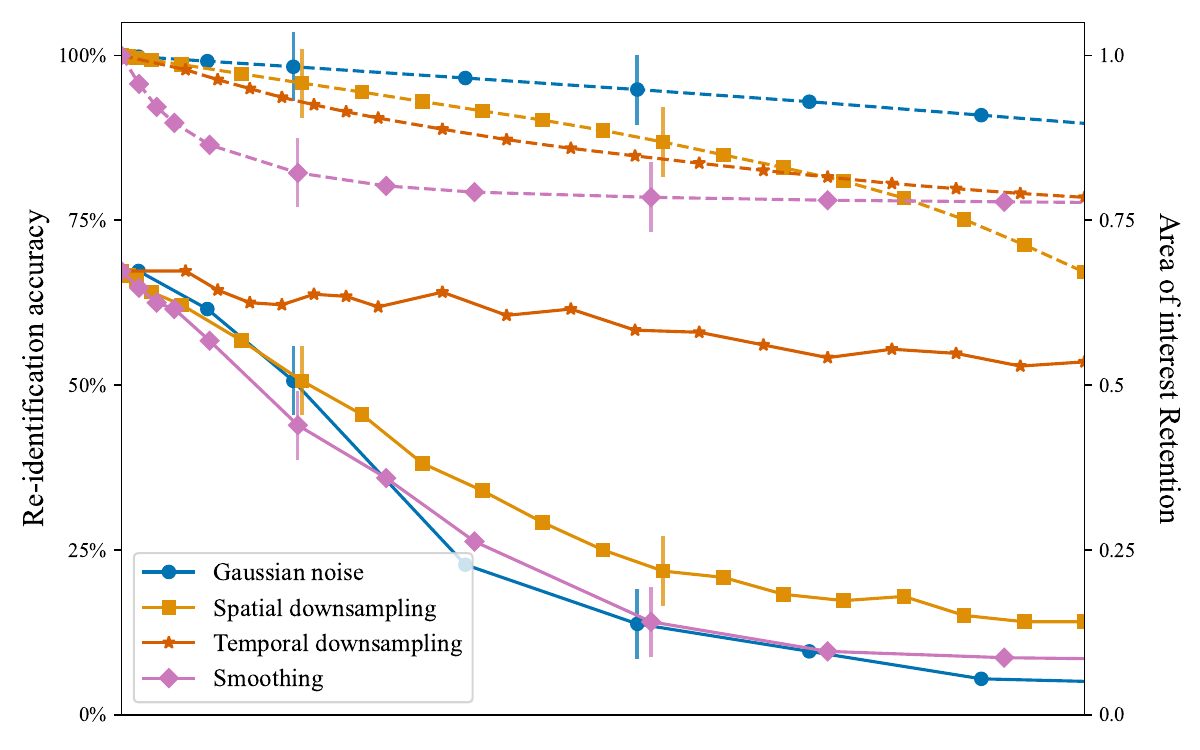}
    \caption{\editnote{Changed color scheme to be consistent with later figures.}Identification accuracy (solid lines) and AOI retention (dotted lines) of privacy mechanisms applied at various strengths.  X axes have been scaled to provide roughly the same privacy falloff so that utility can be directly compared.  Vertical lines indicate the chosen low and high strengths of each mechanism.}
    \label{fig:e1-pu}
\end{figure}

We compute identification accuracies for each mechanism at multiple strengths to define a privacy curve and measure against AOI F1 scores to estimate the anticipated utility trade-off.  We measure Gaussian noise up to $\sigma = 20\degree$, spatial downsampling up to $L = 256$, temporal downsampling up to $K = 30$, and smoothing up to $B = 300$.  The privacy/ utility trade-off can be seen in Figure~\ref{fig:e1-pu}.

Of the previously proposed mechanisms, additive Gaussian noise and spatial downsampling are effective at providing privacy.  The proposed smoothing mechanism also provides privacy, albeit at a higher trade-off in AOI retention.  We have not found temporal downsampling to be an effective privacy mechanism.  It is likely that by retaining $N/K$ real samples, the model can still associate real samples and distinguish users until $K$ reaches a point far beyond usable utility.

We introduced smoothing as a novel mechanism to protect user eye movements.  On our data, smoothing shows potential for privacy protection with similar levels of privacy attained as Gaussian noise and spatial downsampling.  However, AOI retention is lowest for smoothing on the tested dataset, indicating that an application receiving smoothed data would have less reliable accuracy per-sample than other mechanisms.  However, as we will see in Section~\ref{sec:user-centric-utility}, lower AOI retention is not reflective of smoothing's impact to user experience.

%% file: pages/3.table.tex
\begin{table}[t]
\centering
\resizebox{\linewidth}{!}{%
\begin{tabular}{c|r|c|c} 
\toprule
\textbf{Duration} & \multicolumn{1}{c|}{\textbf{Test Data}} & \multicolumn{2}{c}{\textbf{Train Data}} \\ \hline  
& \multicolumn{1}{c|}{} & \begin{tabular}[c]{@{}c@{}}\textbf{Conventional}\\\textbf{Eye Tracking}\end{tabular} & \begin{tabular}[c]{@{}c@{}}\textbf{Constrained}\\\textbf{VR}\end{tabular}  \\ 
\cline{3-4}
\multirow{3}{*}{5 s}
& Conventional Eye Tracking & 66.09\% & 18.04\% \\
& Constrained VR & 27.04\% & 33.87\% \\
& Interactive VR & 22.44\% & 26.92\% \\

\rowcolor[rgb]{0.902,0.902,0.902} & Conventional Eye Tracking & 81.42\% & 32.36\% \\
\rowcolor[rgb]{0.902,0.902,0.902} 10 s & Constrained VR  & 36.00 \%  & 45.14\% \\
\rowcolor[rgb]{0.902,0.902,0.902} & Interactive VR & 19.23\% & 22.76\% \\

 & Conventional Eye Tracking & 90.00\%  & 54.58\%  \\
30 s & Constrained VR & 48.57\%  & 55.42\% \\
 & Interactive VR & 40.06\% & 41.35\%  \\

\rowcolor[rgb]{0.902,0.902,0.902} {\cellcolor[rgb]{0.902,0.902,0.902}}& Conventional Eye Tracking & 90.78\% & 58.57\% \\
\rowcolor[rgb]{0.902,0.902,0.902} {\cellcolor[rgb]{0.902,0.902,0.902}}& Constrained VR & 49.98\% & 55.09\%\\
\rowcolor[rgb]{0.902,0.902,0.902} \multirow{-3}{*}{{\cellcolor[rgb]{0.902,0.902,0.902}}60 s} & Interactive VR & 56.73\% & 54.81\% \\
90 s & Interactive VR & 68.27\% & 67.31\%
\end{tabular}
}
\caption{Average re-identification accuracy using the EKYT architecture~\cite{lohr_eye_2022} trained and evaluated on different data types.  All models are trained at 125Hz.}
\label{tab:tab}
\end{table}

%% file: pages/5user-in-the-loop.tex
\section{Evaluating User-centric Utility}
\label{sec:user-centric-utility}

\begin{figure*}[t]
    \centering
    \includegraphics[width=1\linewidth]{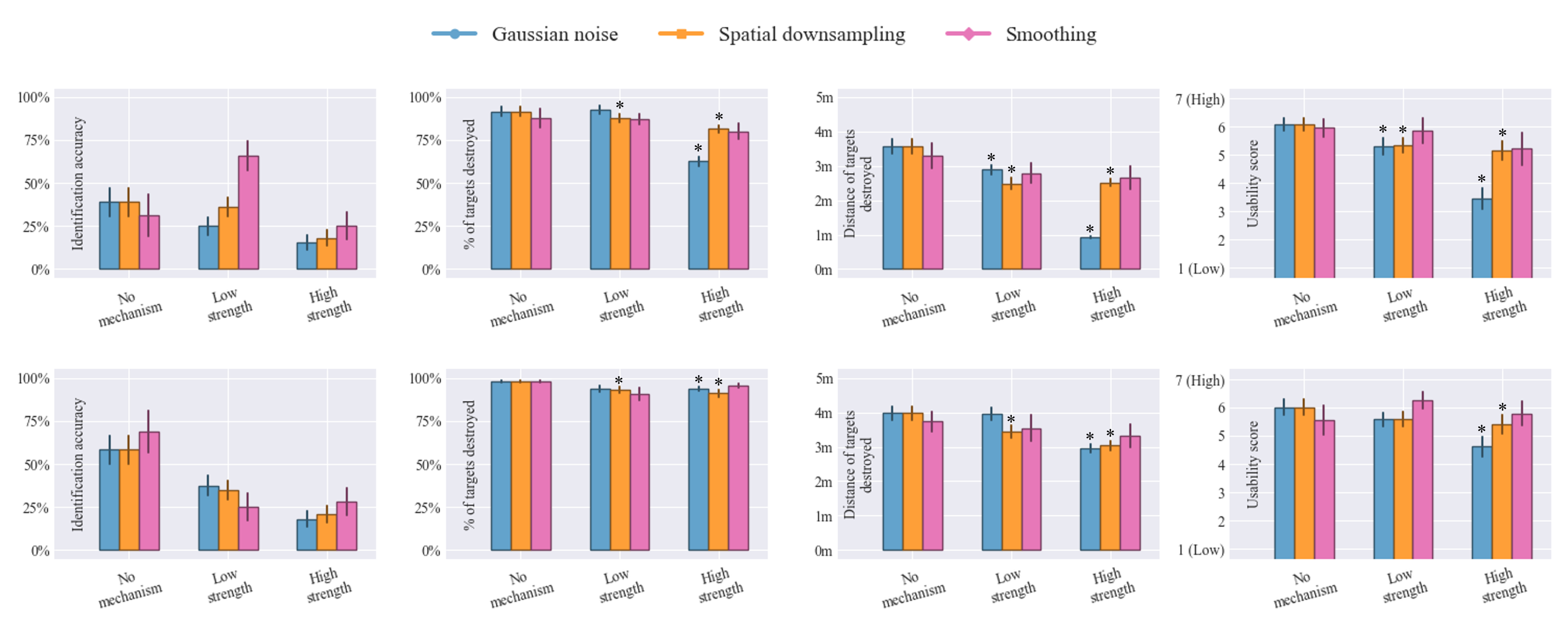}
    \caption{\editnote{Updated this figure to be a barchart with error bars.  \textbf{Reviewer comment: Fig. 3 should be plotted in bar charts with error bars. Without error bars, the current line charts cannot show the distributions of samples. It's hard to assess why two mechanisms with almost the same means but their statistical analysis results, e.g., p-values, are quite different. For example, the percentages of destroyed targets in the three high-strength mechanisms all have very similar means, but only two of them are significantly worse than the baseline (no mechanism). Bar charts with error bars or box plots would present this result much better.}}  Collected metrics of the immersive VR game with gaze controls.  For identification accuracy (first column), lower values indicate more privacy.  For all utility metrics, a higher score indicates higher utility.  Each subplot illustrates no privacy mechanism compared against the low and high strengths of each mechanism.  Wilcoxon signed-rank test significance for utility metrics ($p < 0.05$) are denoted with color-coded asterisks.  \editaddsmall{Vertical lines indicate Standard Error of the Mean (SEM) $= \sigma / \sqrt{N}$.}}
    \label{fig:e2-pu}
\end{figure*}

\begin{figure*}[t]
    \centering
    \includegraphics[width=1\linewidth]{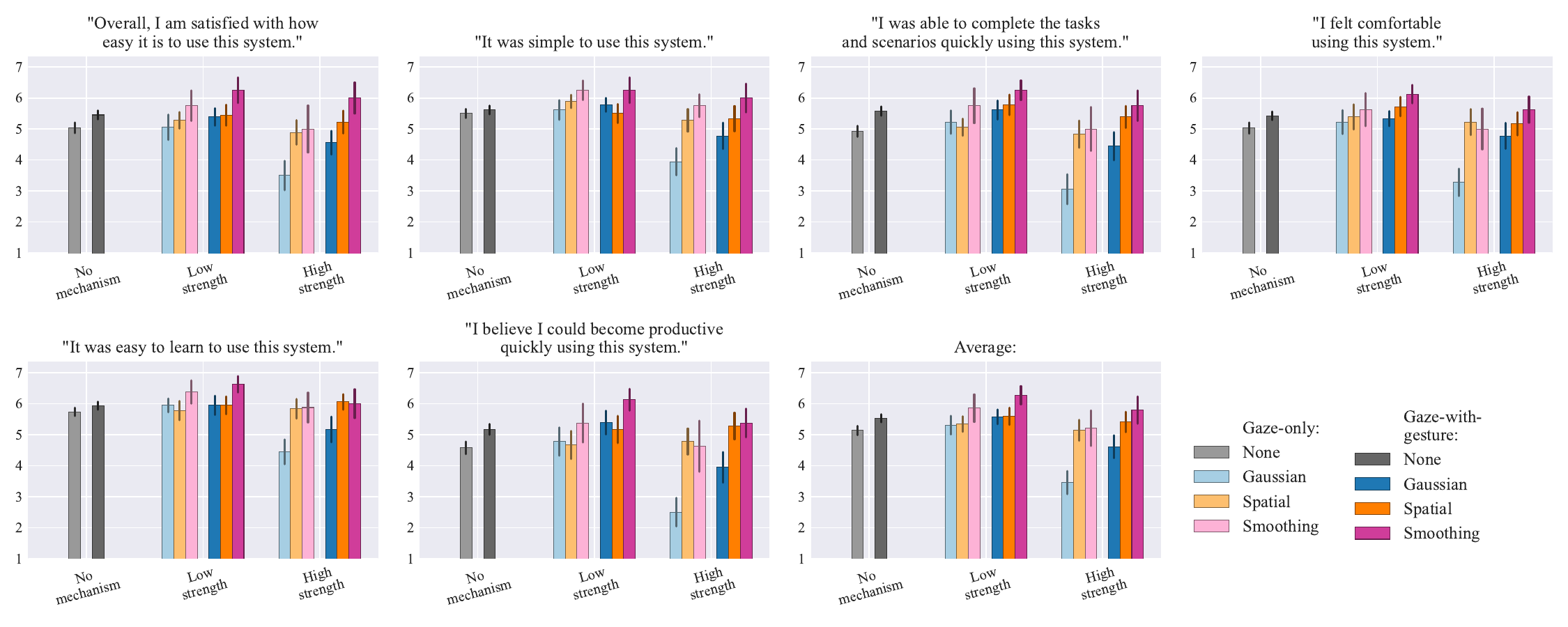}
    \caption{
    \editnote{Changed the outline and error bar colors to be less harsh.  Updated to colorblindness-friendly colors.} 
 Breakdown of PSSUQ SYSUSE scores~\cite{lewis_psychometric_1992} across conditions by individual question.  Vertical lines indicate \editaddsmall{SEM}\editremove{Standard Error of the Mean (SEM) $= \sigma / \sqrt{N}$}.}
    \label{fig:pssuq}
\end{figure*}




The majority of privacy literature surrounding eye tracking movements have considered privacy as a post-process operation.  However, in the context of interactive VR, privacy must be provided \textit{before} being passed to the application; the user subsequently sees the impacts of privatization.  It is important to consider the user and the trade-off in \textit{user-centric utility}, rather than solely relying on data-centric metrics.  In this section, we answer the following research question:

\begin{itemize}[noitemsep, topsep=0pt]
    \item What is the impact to user-centric utility metrics when eye tracking privacy mechanisms are applied live in interactive VR?
\end{itemize}

We investigate user-centric utility through an interactive VR experience which uses eye gaze as the primary control mode.  Before being processed by the application, the eye movement data is processed by the 3 viable privacy mechanisms: Gaussian noise, spatial downsampling, and smoothing.

\subsection{Data Collection Methodology}

We describe the protocol for our collected dataset in which privacy mechanisms were applied in real-time, allowing us to measure user-centric utility after privatization in interactive VR.

\textbf{Participants:}
Survey participants were recruited under IRB approved protocol via several communication channels including word of mouth and electronic mailing list advertisements ($N=18$; 77.78\% male, 22.22\% female).  Eligible participants required normal or corrected-to-normal vision without the use of eye glasses.  The racial-ethnic distribution is 33.33\% White, 22.22\% Indian Asian, 11.11\% Black or African American, 16.67\% Hispanic/Latino, 5.56\% Eastern Asian, 5.56\% Central Asian, and 5.56\% Middle Eastern.  27.78\% of participants were age 18-20 and 66.67\% 21-29.  88.89\% of participants reported some level of experience with VR, and 16.67\% reported some experience with using eye tracking as a control mode.  
The smoothing mechanism was incorporated into the study after 10 participants had undergone a version with only Gaussian noise and spatial downsampling, so for all measures of smoothing, there are $N=8$ participants represented.  This detail is addressed in our results.

\textbf{Procedure:}
Participants played a first person shooter game where they remained in a static position and defeated enemies that periodically spawned and travelled towards the player.  Two types of entities would alternatively spawn in random positions and move towards the player.  Friendly entities served as visual distractors~\cite{komogortsev_fast_2012} and enemy entities served as targets; participants were instructed to destroy targets before they could reach the player.  Entities would spawn randomly from 7 uniformly spaced points on a 90$\degree$ arc spaced 10 meters from the player position.  Participants could see a translucent gaze cursor indicating their current gaze direction.  In trials with privacy mechanisms applied, the cursor illustrated the effective gaze direction after privatization.  See Figure~\ref{fig:teaser}b for a participant's typical view.

Each trial of the game lasted for $\sim$30 seconds.  30 entities (15 targets and 15 distractors) spawned, increasing in speed from 1 m/s to 5 m/s over the duration of the trial.  By increasing difficulty over the course of the trial we can derive useful performance metrics from participants regardless of skill level.  For each experimental condition, participants would undergo two trials of the game sequentially, followed by the Post-Study System Usability System Usefulness subscale (PSSUQ SYSUSE)~\cite{lewis_psychometric_1992}.

We adopted a within-subjects design; all participants underwent every condition.  Privacy mechanism conditions were either no privacy mechanism applied or \{Gaussian noise, spatial downsampling, smoothing\} at \{low, high\} strength, presented in random order.  The privacy mechanism strengths chosen were derived from the re-identification accuracies found in initial analysis the data presented in Section~\ref{sec:e1-results}\footnote{Using preliminary results \editaddsmall{from Section~\ref{sec:e1-results}}, the parameters which initially decreased re-identification accuracy below 40\% and 20\% were chosen for low and high strengths.}.  For Gaussian noise, low $\sigma = 1\degree$ and high $\sigma = 3\degree$.  For spatial, low $L = 48$ and high $L = 144$.  For smoothing, low $B = 50$ and high $B = 150$.

We implement two control modes for facilitating the selection of targets.  The first control mode is \textit{gaze-only}; if the gaze vector from the participant's left eye was consistently within a target's bounds for 500 ms, the target was destroyed.  The other control mode is \textit{gaze-plus-gesture}; participants' gaze vectors indicated the selection of targets.  If, for a given frame, the player gaze vector intersected with a target's AOI and the participant performed a pinch gesture with their left hand, the selected target was destroyed.  

Participants underwent all privacy mechanism conditions in random order with one control mode, then all privacy conditions again in random order with the remaining control mode.  To mitigate an order-effects bias, the order of control mode was counterbalanced among participants.  

\textbf{Validation:}
Data was collected using the Meta Quest Pro~\cite{meta_quest_pro}.  Before undergoing the experiment, each participant underwent the same validation protocol as in Section~\ref{sec:validation}, but this time underwent 3 trials rather than 1 (spatial accuracy error $\mu = 2.78$, $\sigma = 1.45$).  Participants also performed gesture validation.  Again facing a $3 \times 3$ board of targets, participants were instructed to place a cursor at the center of the field of view over the target and make a pinching gesture to confirm their placement (error $\mu = 2.94$, $\sigma = 1.68$).  While gestures were not explicitly calibrated per participant, this served as a primer for participants so that they were familiar with the headset's gesture recognition before undergoing the main tasks.

\subsection{Evaluation Methodology}

We analyze identification accuracies alongside multiple notions of user-centric utility to clearly measure whether the addition of privacy mechanisms have any negative impact to users.  These metrics are:
\begin{itemize}[noitemsep, topsep=0pt]
    \item \textbf{\% of targets destroyed:} This is a simple measure of task performance, conveying on average how many targets were successfully destroyed before reaching the participant.
    \item \textbf{Average distance of targets destroyed:} Also measuring task performance, this metric gives a better estimate of how easy/ challenging the task was for users.
    \item \textbf{Usability score:} The average PSSUQ SYSUSE response, representing users' perceived satisfaction with the given control mode and privacy mechanism combination.
\end{itemize}

\subsection{Results}

We present our results across identification accuracy and the task-specific notions of utility defined above.  Note that when computing identification accuracy, we pair-wise compare the first and second trial of a given mechanism against those of no privacy mechanism to form our query and reference setups.  We test for statistical significance using the Wilcoxon signed-rank test after collapsing the results of all game trials to a single data point per participant and condition, reporting values of $p < 0.05$ as significant.  We compare against the non-privatized trials when testing significance.  Results are visualized in Figure~\ref{fig:e2-pu}.

We first see a lower identification accuracy ($\mu = 48.61\%$) on this dataset than the dataset of Section~\ref{sec:e1-dataset}.  This is to be expected, as session lengths are $\sim$30 seconds rather than 90.  In this new experimental setup, the high strength privacy mechanisms all provide a noticeable level of privacy.

The percentage of targets destroyed and distance targets are destroyed at function as measurements of task performance.  For gaze-only controls, we see a decrease in performance across all privacy mechanisms.  For the \% of targets destroyed, Gaussian ($z=-3.73)$ and spatial ($z=-3.01$) are significantly lower than the baseline at high strengths.  For the average distance, Gaussian ($z=-2.94$) and spatial ($z=-3.72$) are significant at low strength, and Gaussian ($z=-3.72$) and spatial ($z=-3.1$) are significant at high strength.  It is more difficult for the application to measure continuous fixations with the perturbed data, so performance decreases.  Interestingly, we see less of an effect when participants used gaze-plus-gesture controls.  For \% of targets destroyed, spatial ($z=-2.57$) is significant at low strength, and Gaussian ($z=-2.98$) and spatial ($z=-3.06$) are significant at high strength.  For average distance, spatial ($z=-2.63$) is significant at low strength and Gaussian ($z=-3.03$) and spatial ($z=-2.98$) are significant at the high strength.  Generally, the average percentage of targets destroyed remains high, though the average distance decreases.  This indicates that the task became slightly more difficult but remained trivial to complete.  Across both task performance metrics, we see Gaussian noise utility decrease at a high rate compared to the other mechanisms.  

Focusing on usability, we see noticeable differences in average performance \editaddsmall{(illustrated in Figure~\ref{fig:pssuq})}.  For gaze-only controls, Gaussian ($z=-2.82$) and spatial ($z=-2.74$) are significant at low strength and Gaussian ($z=-3.64$) and spatial ($z=-2.59$) are significant at high strength.  For gaze-plus-gesture controls, Gaussian ($z=-3.59$) and spatial ($z=-2.1$) are significant only at high strength.  Gaussian noise again impacts utility at a higher rate than other mechanisms.  Interestingly, smoothing seems to slightly increase usability when applied at a low strength.  There could be a low amount of noise present in the eye tracker's raw data stream which smoothing corrects.  Yet, there is not sufficient statistical evidence to prove this claim.  For both non-Gaussian mechanisms, overall usability remains high even after applying the high strength variant of the mechanisms.  \editremove{Figure~\ref{fig:pssuq} illustrates user responses across individual questions.}

\if\revisionmode1
\hl{Below are results of a 3-way ANOVA on average usability score:}
{ \small \color{blue}
\begin{verbatim}
    3-way ANOVA on PSSUQ average (N = 8):
=============================================================
                                F Value Num DF  Den DF Pr > F
-------------------------------------------------------------
mechanism                       10.6902 2.0000 14.0000 0.0015
strength                         6.8161 2.0000 14.0000 0.0086
control_mode                     0.2348 1.0000  7.0000 0.6428
mechanism:strength               7.3979 4.0000 28.0000 0.0003
mechanism:control_mode           4.2953 2.0000 14.0000 0.0351
strength:control_mode            2.4520 2.0000 14.0000 0.1222
mechanism:strength:control_mode  0.7953 4.0000 28.0000 0.5383
=============================================================
\end{verbatim}
}
\fi

\editadd{[R1] As there are multiple factors in the results of Fig. 3, I wonder whether two-way or three-way ANOVA has been conducted. There might be interactions between mechanism and strength, or between control modes and mechanisms.}{\editnote{We ran a 3-way ANOVA.  The literal output is pasted above as a review note, and the following paragraph covers our results in the text.}  We conducted a three-way ANOVA on the participants which underwent all privacy mechanisms ($N = 8$) to examine the effect of privacy mechanism, strength, and type of control mode on the average PSSUQ response. We find the main effects of mechanism and strength to be significant ($p < 0.01$).  There are significant interactions between mechanism and strength ($F(4, 28) = 7.3979, p=0.0003$) and between mechanism and control mode ($F(2, 14) = 4.2953, p=0.0351$).  These findings evidence that the different control modes could be affected by the application of privacy mechanisms disproportionately.}

\editadd{[R1] In Fig. 3, gaze-only control condition, the identification accuracy in the low-strength smoothing mechanism is higher (lower privacy) than that without any mechanism. Any explanation for this failure to preserve privacy?}{\editnote{We believe this outlier is due to our set of smoothing individuals being smaller, the results are more volatile.  Reworded this section to better clarify this.}  Note that because smoothing was implemented midway through data collection, statistical tests of smoothing consider $N=8$ participants, thus there is less statistical power than other conditions.  However, across utility metrics, smoothing appears comparable to spatial downsampling across all categories with comparable variance.}
\editremove{We note that due to smoothing being introduced midway through participant recruitment, identification accuracy should not be compared against the other methods and there is less statistical power present for utility comparison against the baseline condition. with the data available, smoothing appears equivalent or better than spatial downsampling across all utility categories.}

%% file: pages/6threat-scenarios.tex
\section{Anticipated Adversarial Threats}
\label{sec:threats}




So far, we have presented an analysis of the identification potential in interactive VR and presented privacy solutions which can mitigate the risk of identification while retaining user-centric utility.  However, this is only the first step.  We must also consider the robustness of any privacy solution, as a dedicated adversary will make efforts to counteract and nullify any privacy-preserving operation.

The eye tracking community has explored formal privacy guarantees such as differential privacy (DP)~\cite{liu_differential_2019, bozkir_differential_2021, steil_privacy-aware_2019}, k-anonymity or plausible deniability~\cite{david-john_for_2022, david-john_privacy-preserving_2023}.  However, as the privacy guarantee is tied to high level features and applied to full collections of users, these methods are not suitable for sample level eye tracking data being privatized in real time.  Kal$\varepsilon$ido has developed a sample-level method with DP guarantees~\cite{li_kalido_2021}, but has a high overhead (8ms), and \>15-20ms overall latency in VR can introduce sickness and nausea~\cite{abras_latency_2012}.  Thus, privacy mechanisms proposed for live gaze-based interactions must be proactively evaluated against adversarial threats.  In this section, we answer the following research question:

\begin{itemize}[noitemsep, topsep=0pt]
    \item Are the proposed privacy mechanisms robust against malicious adversaries?
\end{itemize}

We address the robustness of our proposed mechanisms under three realistic threat models in which adversaries have varying levels of domain information.  We then define and evaluate an example attack under each threat model.  For this analysis, we use the larger dataset defined in Section~\ref{sec:e1-dataset} and implement privacy mechanisms at their high strength.

\subsection{Threat Models}

We explore multiple threats to eye tracking re-identification that we expect to become plausible in the next decade as eye tracking technology and VR become more mainstream.  Our threat models are organized according to the information or resources that the adversary has access to; as an adversary becomes more informed of the privacy mechanism, they become increasingly able to counteract the privacy efforts.  We conceptualize an example scenario and attack for each threat model.

In all cases, we make the assumption that the adversary's goal is to obtain the identity of an acquired query gaze stream.  There may be sensitive information connected to either the query gaze stream or existing dataset records, and a successful attack can link the user's identity or quasi-identifier(s) to the sensitive information.  These sensitive attributes could be concrete records, such as health information or group membership, or could be implicit knowledge embedded within in the actual gaze stream (such as personality~\cite{berkovsky_detecting_2019}, age~\cite{zhang_how_2018} or gender~\cite{sammaknejad_gender_2017}).  

\textbf{Black-box Access:}
In this scenario, an adversary has acquired a privatized gaze stream but has no knowledge regarding the mechanism applied.  The adversary can attempt to query the privatized record against non-privatized records sourced from elsewhere, such as public datasets.  Before querying, it is possible for the adversary to perform a filtering operation in an attempt to render the privacy mechanism ineffective~\cite{papadimitriou_time_2007}.

\begin{indentparagraph}
    \textit{A malicious VR gaming application (the adversary) records eye tracking data that has been securely privatized by a user's VR hardware.  By enabling the posting of high scores to social media, the adversary learns the user's identity.  The adversary can then query against released datasets with associated medical diagnosis (autism, alzheimer's, depression, etc.), attempting to verify the user's membership in the dataset.  If successful, the adversary has a platform to perform fraud or blackmail.}
\end{indentparagraph}

\textbf{Black-box Access with Exemplars:}
Similar to the above scenario, the adversary does not have knowledge about the implementation of the applied mechanism.  However, the adversary has access to a large number of privatized records, possibly paired with a number of non-privatized records.  From here, the adversary could attempt to approximate the mechanism, or perform regression analysis to learn an inverse function of the mechanism.

\begin{indentparagraph}
    \textit{An adversary has a new VR headset which only releases privatized gaze data streams and an older model which releases raw gaze vectors.  They recruit a number of users to perform the same tasks while wearing both headsets.  When they have a sufficient amount of data, they model a function to invert the privacy mechanism, increasing the chance of re-identification.  From this point on, they can apply that function to other records collected through the same hardware.}
\end{indentparagraph}

\textbf{White-box Access:}
In this scenario, the adversary knows the exact implementation of the mechanism that has been applied.  This could be learned from data leaks of design documents or code~\cite{mccormick_data_2008}, or by guessing simpler mechanisms by observing a sufficient number of samples and approximating parameters.  

\begin{indentparagraph}
    \textit{An insider of a VR hardware company posts the confidential privacy algorithm to an online forum.  From there, any adversary who obtains the algorithm can attempt to leverage that knowledge against privatized data records obtained from the device.}
\end{indentparagraph}

\subsection{Evaluation Methodology}

We illustrate the risk of each defined threat scenario with toy example attacks, simulating an adversary with the corresponding amount of knowledge.  These attacks are not optimized or exhaustive, but illustrate the additional risks of data leakage that have not been widely considered in eye movement privacy literature.  

\textbf{Wavelet Denoising:}
In the black-box threat scenario, an adversary can perform an uninformed filtering attack to attempt to nullify the effectiveness of the privacy mechanism.  Time series perturbations that are implemented on independent samples are prone to filtering attacks, which can vastly reduce uncertainty if the pattern can be filtered out~\cite{papadimitriou_time_2007}.

We illustrate this by applying a wavelet denoising filter~\cite{donoho_adapting_1995} over the privatized data stream.  The implementation\footnote{\url{https://scikit-image.org/docs/stable/api/skimage.restoration.html\#skimage.restoration.denoise_wavelet}} assumes a level of noise and estimates $\sigma$ automatically, requiring no knowledge of the mechanism at hand.    

\textbf{CNN Data Regression:}
In the black-box with exemplars scenario, the adversary has acquired a number of data samples with and without privatization.  Data driven approaches could be implemented in an attempt to approximate the privacy mechanism or to directly approximate an inverse function.

We illustrate this concept with a simple convolutional neural network (CNN) which trains to reconstruct input privatized data streams back to the original data streams.  Our implementation inputs and outputs 5 seconds of data.  The model consists of 4 [1D Convolution, 1D Batch normalization, Tanh] blocks.  For each condition, we train a model on 50\% of data then evaluate on 50\%, repeat the process with reverse train-test splits, then report the averaged accuracy.

\textbf{Mechanism Applied to Reference Data:}
In the white-box scenario, the adversary knows the implementation details at hand.  Obviously, if the mechanism is deterministic it becomes possible to reconstruct the original data by reversing the process.  However, in stochastic mechanisms a level of uncertainty remains.  

We investigate the white-box scenario by leveraging the adversary's inside knowledge to apply the same operation to the reference set.  If the adversary does not know the query identity, but does know the identities of non-privatized reference records, the adversary could apply the privacy mechanism's algorithm across the board for a more equal comparison.

\subsection{Results}
\label{sec:threat-model-results}

We report the results of our toy examples for each threat scenario in Table~\ref{tab:threat-scenarios}.  We find that Gaussian noise is vulnerable to attacks across all threat scenarios.  Spatial downsampling and smoothing are not vulnerable to the black-box scenario’s filtering attack, but all mechanisms are vulnerable to a degree to each scenario in which adversaries have additional knowledge.  However, these mechanisms significantly increase the amount of knowledge and effort required to successfully re-identify users.  Presumably, a white-box with exemplars threat scenario would have access to attacks that are even more successful.

In the black-box with exemplars scenario, both Gaussian and smoothing identification accuracies after CNN regression are higher than the original re-identification rate.  It is possible that the CNN's inverse approximation accentuated some important features from the original data stream, potentially making the undone data streams slightly more identifiable.

Smoothing is more vulnerable than spatial in the black-box with exemplars scenario, being brought above the original identification accuracy.  This can be attributed to our smoothing implementation being a fully deterministic process; as a result, the original signal can be fully reconstructed if the first real data value and exact buffer size is known.  However, there are a number of small optimizations which could be made to the smoothing process, such as non-uniformly initializing the buffer or adding random variance to the impact of each weight.

\begin{table}[t]
\centering
\resizebox{\linewidth}{!}{%
\begin{tabular}{r|cccc}
\textbf{Mechanism} & \begin{tabular}[c]{@{}c@{}}\textbf{ID}\\\textbf{Accuracy}\end{tabular} & \textbf{Black-box} & \begin{tabular}[c]{@{}c@{}}\textbf{Black-box}\\\textbf{with exemplars}\end{tabular} & \textbf{White-box}  \\ 
\hline
None               & 67.31\%                                                                & /               & /                                                                                & /                     \\
Gaussian           & 14.1\%                                                                & 63.14\%         & 68.39\%                                                                          & 64.74\%                  \\
Spatial            & 21.79\%                                                                & 21.79\%         & 61.47\%                                                                          & 58.33\%               \\
Smoothing          & 14.1\%                                                                & 14.1\%         & 69.49\%                                                                          & 55.13\%              
\end{tabular}
}
\caption{Identification accuracies before and after performing attacks on privatized data across three threat scenarios.}
\label{tab:threat-scenarios}
\end{table}

%% file: pages/7discussion.tex
\section{Discussion}

We find that there is some risk of re-identification from eye movements collected in VR applications.  On 125Hz data and using models trained and/or evaluated on VR data, we report accuracies of up to 33.87\%@5s, 58.57\%@60s, and 68.27\%@90s.  Our upper measure of 90 seconds approaches reliable identification; yet, commercial VR applications such as games or virtual training scenarios can have much longer sessions.  Countermeasures should be designed with this in mind.


We find that across all user-centric metrics of utility, Gaussian noise is lower than the other evaluated mechanisms.  Conversely, smoothing has the lowest data-centric AOI retention but highest user-centric utility.  This finding highlights that when developing privacy mechanisms for interactive VR, it is critical to be user-centric.  The findings from Section~\ref{sec:e1-results} and prior work~\cite{david-john_privacy-preserving_2021} both would suggest Gaussian noise to be the best of evaluated mechanisms, but our analysis suggests that Gaussian noise should be rejected from a user experience standpoint.  Spatial downsampling and smoothing are viable as privacy mechanisms in interactive contexts, but more work should be done to increase robustness against knowledgeable adversaries.

\textbf{Broader Impacts:}
This work extends the discussion of privacy in eye tracking and VR by placing a focus on the real-world implications when applying privacy mechanisms to future applications.  A large emphasis should be placed on user-centric notions of utility for gaze-based interaction applications, rather than only evaluating data-centric utility.  If user experience is compromised, users will not be willing to engage in VR experiences in the first place.  When evaluating privacy mechanisms, on top of the simplest case where re-identification accuracies are compared before and after privatization, researchers should test against more challenging threat models grounded in real world scenarios.

\editadd{[R1] Also related to the 1st issue, for this privacy-preserving problem of eye tracking data, are there any issues that are uniquely related to VR? It seems that the proposed approach and study can be used in other non-VR applications, e.g., web browsing, and computer games.}{The evaluated privacy mechanisms could be applied to a larger set of eye tracking applications, such as augmented reality (AR) settings or to webcam eye trackers.  We believe VR technology to be the most pressing use case currently; eye gaze is a promising input device showcasing unique interactions and enabling critical optimizations such as foveated rendering, but users should not have to choose between these features or their own privacy.}

\textbf{Limitations:}
Our user-centric evaluation relies on a single interactive VR dataset. This dataset includes gaze-only selection and gaze-with-gesture selection.  This does not represent the full diversity of gaze-based interactions in VR, each of which may have their own nuances and thresholds for what is a reasonable level of utility traded for privacy gained.  For example, consider an eye tracking-enabled competitive gaming context.  Users in that context are unlikely to accept any privacy mechanism that compromises performance.

The analysis in Section~\ref{sec:threats} highlights the importance of robustness; however, our list of threat models are not exhaustive.  There are a large number of adversaries and attack methods yet to be considered.

\textbf{Future Work:}
In Section~\ref{sec:threat-model-results}, we mentioned some improvements to smoothing that could introduce randomness.  These improvements could increase smoothing's robustness while retaining a level of usability.  Additionally, composition of simple operations may yield better mechanisms (passing spatial downsampled values into the smoothing buffer, for example, could be explored).

A methodology for further improved mechanisms could be to introduce temporal perturbations alongside sample-level perturbations.  As the operation needs to be possible in real-time, it is difficult to introduce temporal inconsistencies without some form of delay.  Potential avenues to explore would be to leverage context of objects from the scene~\cite{erdemir_privacy-aware_2021} to modulate dwell times and durations between fixations, or to jointly perform privatization and gaze prediction to offset any delays.

There is a strong correlation between the ability to be identified and the amount of data available.  An evaluation could be done on long data sessions (say, 30+ minutes continuously) to further quantify this risk at durations expected in VR applications. One mitigation would be to apply a random mechanism from a selection of viable mechanisms every 1 or 2 minutes, making the full session unreliable for queries. 

\editadd{[R1] How do the privacy-preserving mechanisms affect spatial perception, cybersickness, or multisensory perception? It might be interesting to consider measuring workload or cybersickness in the experiments.}{It is currently unclear what potential impacts our mechanisms would impose on other user-experience enhancing research focused on eye tracking in VR.  Future work could explore concepts such as spatial perception~\cite{xiang_work--progresspreliminary_2022}, cybersickness~\cite{islam_cybersickness_2021,lopes_eye_2020}, and multisensory perception~\cite{marucci_impact_2021} and the level of impact that eye tracking privacy mechanisms would have on these metrics.}

%% file: pages/8conclusion.tex
\section{Conclusion}

We analyzed the re-identification risk associated with eye tracking-enabled interactive VR applications and evaluated multiple privacy mechanisms which could serve as potential solutions for mitigating risk.  In this work, a large emphasis was placed on user-centric notions of utility.  While prior work has focused on data-centric utility, there is a necessary shift towards user-first design for applications where the user directly interfaces with the eye tracking functionality.  We also further investigate real-world feasibility, modeling multiple threat scenarios where adversaries can attempt to counteract privacy efforts.

This work shows that there is re-identification risk associated with eye tracking in interactive VR, though the risk is less prominent than prior analyses with conventional eye tracking systems.  Of the mechanisms evaluated, we found spatial down sampling and smoothing to be viable for practical applications.  These mechanisms provide privacy while retaining high subjective usability and reasonable task performance, yet each mechanism is vulnerable against highly informed adversaries.

We hope that this work will aid further research regarding privacy protection in interactive VR applications.  By placing a focus on user-centric utility and highlighting real world threat scenarios, we provide a methodology for the analysis of privacy mechanisms that puts the user first.

%% file: pages/zappendix.tex
\appendix
\section*{Appendix}

\editnote{To address reviewer 3's comments on the runtime analysis of our mechanisms and data collection details, we have added an appendix.  According to TVGC guidelines available here (https://ieeecs-media.computer.org/tc-media/sites/49/2023/03/14151934/vgtc\_mar23.pdf), appendices will not be included in the conference/ journal publications, but we will host a preprint upon full acceptance with this appendix added.}

\section{Description of Privacy Mechanisms}

\begin{figure}[h]
    \centering
    \includegraphics[width=1\linewidth]{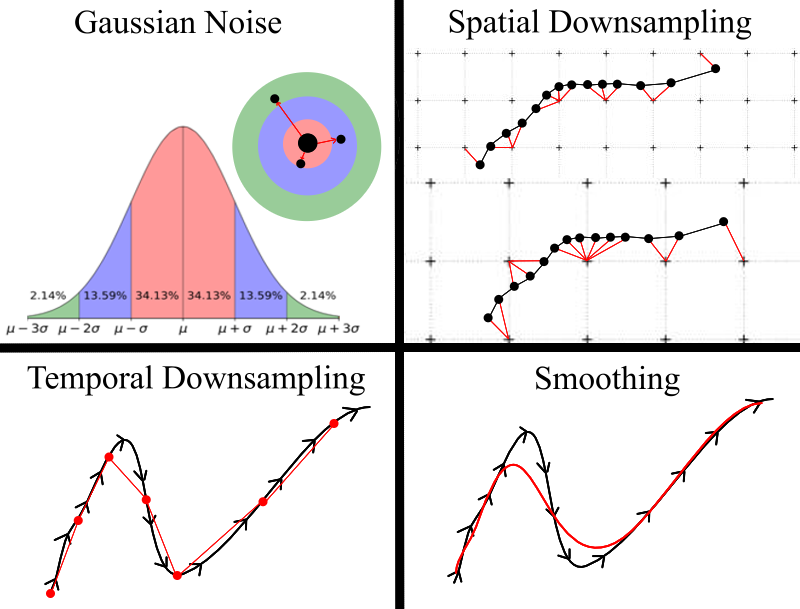}
    \caption{Visualizations of the privacy mechanisms implemented in our experiments.}
    \label{fig:vizzy}
\end{figure}

Here we provide illustrations and short descriptions of the concepts of our privacy mechanisms.  See Figure~\ref{fig:vizzy} for illustrations.  Gaussian noise offsets each gaze data point independently at each frame by drawing upon a Gaussian distribution.  As the noise sampling is independent, no temporal patterns can be discerned to identify individuals; however, the mechanism is susceptible to filtering to recover the original signal.  Spatial downsampling maps the continuous range of gaze values to an equirectangular grid of discrete points.  The true gaze angle is mapped to the closest discrete value at every frame.  Temporal downsampling effectively decreases the sampling rate, copying the true value of a frame into the next $N$ frames.  Smoothing applies a linear weighted moving average to a range of gaze values.  The weighting gives recent frames a higher weight than less recent frames.  The result appears to have a slight delay to the user, but by fixating on an object, the smoothed gaze value quickly reaches the intended point.

\section{Runtime Analysis of Privacy Mechanisms}

We provide a simple performance analysis of each of our mechanisms, measuring the impact on device memory and on execution time.  We measure the impact on device memory by reporting the average memory across trials within each privacy mechanism.  We report the total reserved memory by the application and the system memory reported by Unity3D's Memory Profiler\footnote{\url{https://docs.unity3d.com/Manual/ProfilerMemory.html}}.  To measure execution time, we report the average frames per seconds of each trial for each privacy mechanism.  

We see in Table~\ref{tab:runtime} that these mechanisms have little to no impact to performance.  Runtime is not impacted, and system memory increases only slightly on average.  We do see a more noticeable increase in memory usage with smoothing, as it is the only mechanism which stores a continuous array of past gaze samples needed to compute the current gaze sample.  Overall, the mechanisms evaluated are quick to compute and provide negligible performance overhead.  

In our experiments, we process gaze samples provided by the Oculus SDK and apply privacy mechanisms before utilizing the gaze samples in the application.  In a real-world deployment, these mechanisms could be securely implemented on VR HMDs before passing gaze samples to applications.  Figure~\ref{fig:teaser} illustrates the general eye tracking pipeline for VR headsets.  Along with other operations taken to model and process gaze vectors, these privacy mechanisms can be applied on the device securely, then privatized gaze vectors can be provided to potentially untrustworthy applications opened by the VR user.  With current mechanisms' low overhead, these could be implemented on headset software with little performance impact.  Hardware-accelerated implementations could also be explored, and may be more necessary as privacy mechanisms become more complex.

\begin{table}[t]
\centering
\refstepcounter{table}
\resizebox{\linewidth}{!}{%
\begin{tabular}{l|l|l}
\textbf{Mechanism}   & \multicolumn{1}{c|}{\textbf{System Memory (kb)}} & \multicolumn{1}{c}{\textbf{Runtime (FPS)}}  \\ 
\hline
No Mechanism         & 311825 $\pm$ 12443                                  & 70.94 $\pm$ 0.076                              \\
Gaussian Noise       & 318181 $\pm$ 12490                                  & 70.95 $\pm$ 0.083                              \\
Spatial Downsampling & 316606 $\pm$ 13338                                  & 70.94 $\pm$ 0.076                              \\
Smoothing            & 322041 $\pm$ 4818                                   & 70.94 $\pm$ 0.080                             
\end{tabular}
}
\label{tab:runtime}
\caption{Performance analysis of privacy mechanisms on our prototype Unity environment.}
\end{table}

\section{Additional Data Collection Details}

\editadd{[Meta Review] Additional implementation details should be added in relation to the data logger and number of recordings in the final dataset. }{\editnote{In addition to linking our dataset in the main text, we expand upon data collection here.}}

The dataset collected for this publication is available at \url{https://doi.org/10.5281/zenodo.10475455}.  In this section we discuss the implementation of our data logging process and the data collected.

The experiments are implemented in Unity3D using the Oculus SDK to interface with the Meta Quest Pro headsets used for data collection.  At every frame, we query the SDK for gaze samples, process these gaze samples using the current privacy mechanism, then pass the privatized gaze sample for use by the rest of the application.

We log data at every visual frame.  Because the eye gaze provided by the Oculus SDK is only available at every frame's update, gaze data is collected at the application's frame rate.  Data was logged locally to the VR headset's internal storage, then moved after each participant's session.

In \texttt{headset\_data.csv} we log frames, timestamps, active trial conditions, and the position and rotations of the headset, hands, and eyes at every frame.  We also log the non-privatized rotations of the eyes in Experiment 2 for possible comparison.  \texttt{event\_data.csv} contains experiment-relevant events, such as area of interest intersection info, the start and end of trial periods, and the completion of tasks, such as completing a sandwich in Experiment 1 and destroying an enemy in Experiment 2.  \texttt{survey\_data.csv} reports user responses to the PSSUQ questionnaire after each condition.  The dataset also included processed files, which contain streamlined information on conditions, frame, and timestamp and the localized rotations of the eyes.  This processed information is trimmed to only contain frames in which the experiment was active.

In Experiment 1, each participant underwent 4 identical trials of 90 seconds, yielding 9360 seconds of active trial data for $N = 26$ participants.  In Experiment 2, each participant underwent 2 trials of 30 seconds for each privacy mechanism, strength, and control mode pairing, yielding 28 trials for participants which underwent smoothing and 20 for those who did not, yielding 13,200 seconds of active trial data for $N = 18$ participants.